\documentclass[prx,aps,amssymb,twocolumn,noshowpacs,hyperref]{revtex4-1}
\usepackage{color}
\usepackage{graphicx}
\usepackage{hyperref}
\usepackage{amsmath}
\usepackage{latexsym}
\usepackage{textcomp}
\usepackage{amssymb}
\usepackage{mathrsfs}
\usepackage{mathtools}
\usepackage{layout}
\usepackage{verbatim}
\usepackage{multirow}
\usepackage{bm}
\usepackage{amsfonts,epsfig}
\usepackage{lineno}
\usepackage{textcomp}
\usepackage[flushleft]{threeparttable}

\begin{document}
\title{Barium-based Rydberg atom quantum technologies with long Rydberg coherence}
\date{\today}
\author{Xiao-Feng Shi}
\affiliation{Center for Theoretical Physics and School of Physics and Optoelectronic Engineering, Hainan University, Haikou 570228, China}

\begin{abstract}
The short Doppler-limited coherence time of the laser-excited Rydberg state, usually orders of magnitude shorter than the lifetime of the Rydberg state, hinders the Rydberg-mediated quantum technologies. Here, we show that a 649~nm~$-$~658~nm two-photon excitation of the $6sng~^1G_4$ Rydberg state from a long-lived d-orbital clock state of barium can be achieved with a two-photon wavevector that is tiny, which effectively removes the Doppler-limited decoherence. Moreover, the $6sng~^1G_4$ Rydberg state has strong dipole-dipole interaction due to small F\"{o}rster defect with nearby Rydberg states and possesses long radiative lifetime. These can benefit quantum computing based on individually trapped neutral atoms, and can enable long-lived Rydberg polaritons in atomic media, which brings fresh opportunities in all-optical quantum information.

\end{abstract}
\maketitle
\section{Introduction}
Rydberg interactions between neutral atoms~\cite{Gallagh2005} can be used for studying quantum information~\cite{PhysRevLett.85.2208,Lukin2001,Saffman2010,Saffman2016,Saffman2019nsr,Shi2021qst} and quantum many-body physics~\cite{Browaeys2020,Adams2020,Morgado2021,Wu2021,Kaufman2021}, usually with atomic qubits trapped in optical microtraps~\cite{Wilk2010,Isenhower2010,Zhang2010,Maller2015,Jau2015,Zeng2017,Levine2018,Picken2018,Levine2019,Graham2019,Jo2019,Fu2022,McDonnell2022,Bluvstein2022,Graham2022,Evered2023} or with photonic qubits hosted in atomic media~\cite{Dudin2012,Peyronel2012,Firstenberg2013,Baur2014,Tiarks2014,Tiarks2018,PhysRevLett.127.063604}. A stumbling block in these studies is the short ground-Rydberg coherence time $T_{2,\text{\tiny gR}}$. The Rydberg Doppler dephasing time $T_{2,\text{\tiny D}}$ is the major contribution to the ground-Rydberg coherence time $T_{2,\text{\tiny gR}}$ compared to Rydberg radiative decay, atom position variations, and laser noise~\cite{Graham2019}. $T_{2,\text{\tiny D}}$ is inversely proportional to $\Bbbk \sqrt{T_{\text{eff}}}$~\cite{Wilk2010,Saffman2011,Jenkins2012,PhysRevA.101.013421} with $\Bbbk$ the wavevector of the Rydberg excitation and $T_{\text{eff}}$ the temperature characterizing the motional state of the atom. When a two-photon Rydberg excitation is used with two lasers of wavelengths $k_1$ and $k_2$, the smallest possible $\Bbbk$ is $|k_1-k_2|$ when the two lasers counter propagate. In Rydberg excitation of alkali-metal atoms like rubidium and cesium, the smallest two-photon $\Bbbk$ is about $5.0~\mu m^{-1}$ with two counter-propagating lasers of 852~nm and 509~nm for cesium~\cite{Shi2020prapplied}, still leading to $T_{2,\text{\tiny D}}$ which is orders of magnitude smaller than the radiative lifetime of Rydberg states.

The fast ground-Rydberg decoherence hinders Rydberg-mediated quantum information based on single-atom qubits. For example, Ref.~\cite{Evered2023} showed that the first two sources of their gate infidelity are Rydberg $T_{2}^\ast$ and $T_1$, where $T_{2}^\ast$ is identical to $T_{2,\text{\tiny gR}}$. Recently, methods by high-power UV Rydberg excitation of $^{88}$Sr also demonstrated high-fidelity CZ gate~\cite{Finkelstein2024,tsai2024fid}. However, high-power UV lasers are challenging to be realized in parallel large-scale implementations, and though with high laser power their gate error from atom motion is negligible, Ref.~\cite{tsai2024fid} showed that error from $T_1$ and atom motion shoot up quickly, with more contribution from atom motion than from $T_1$ when the Rydberg Rabi frequency decreases beyond $2\pi\times1$~MHz. The effect of atomic motion is dominated by the shot-to-shot Doppler-induced detuning~\cite{tsai2024fid}, which is an alternative term for Doppler dephasing~\cite{Wilk2010}.

In quantum nonlinear optics based on Rydberg atoms, experimentally measured $T_{2,\text{\tiny gR}}$ was on the order of microsecond and shown to be dominated by $T_{2,\text{\tiny D}}$~\cite{Dudin2012,Peyronel2012,Maxwell2013,Tiarks2018,PhysRevLett.131.133001}. This makes cooling atoms to ultra low temperatures necessary since storage of single photons in atomic media is usually required in Rydberg-mediated quantum information processing~\cite{Thompson2017,Tiarks2018}. Alternatively, state mapping between nearby Rydberg states can mitigate motional dephasing in Rydberg polaritons~\cite{shilu2025,Jiao2024}.

\begin{table*}[ht]
     \begin{threeparttable}
  \centering
  \begin{tabular}{|c|c|c|c|c|}
    \hline
 Atom & Rydberg excitation with smallest $\Bbbk$~\tnote{a} &  $ \Bbbk$ & $T_{2,\text{\tiny D}}$ & Lifetime (at 300~K)   \\
     \hline $^{87}$Rb  & $5s~^2S_{\frac{1}{2}} \xleftrightarrow[]{780~\text{nm}} 5p~^2P_{\frac{3}{2}}^{\text{o}} \xleftrightarrow[]{ 479~\text{nm}} ns~^2S_{\frac{1}{2}} $ & $5.1~\mu m^{-1}$ &  20~$\text{ns}\sqrt{\text{K}}/\sqrt{T_{\text{eff}}}    $&  0.19~ms\\
   \hline
 $^{133}$Cs & $6s~^2S_{\frac{1}{2}} \xleftrightarrow[]{852~\text{nm} } 6p~^2P_{\frac{3}{2}}^{\text{o}} \xleftrightarrow[]{509~\text{nm} } ns~^2S_{\frac{1}{2}} $ & $5.0~\mu m^{-1}$ &  25~$\text{ns}\sqrt{\text{K}}/\sqrt{T_{\text{eff}}}$ & 0.18~ms  \\
   \hline
   $^{138}$Ba & $6s5d~^1D_2 \xleftrightarrow[]{649~\text{nm} } 5d6p~^1F_3^{\text{o}} \xleftrightarrow[]{658~\text{nm}} 6sng~^1G_4  $ & $0.14~\mu m^{-1}$  & 930~$\text{ns}\sqrt{\text{K}}/\sqrt{T_{\text{eff}}}$ & 2.3~ms \\
   \hline
  \end{tabular}
  \caption{Comparison of the coherence times of two-photon excited Rydberg states of $^{87}$Rb, $^{133}$Cs, and $^{138}$Ba by taking the transitions with smallest $\Bbbk$ for each atomic species, where $T_{2,\text{\tiny D}}$ is estimated by Eq.~(\ref{eq-T2}). The radiative lifetimes at room temperature for rubidium and cesium by Ref.~\cite{Beterov2009} and that for barium calculated in Sec.~\ref{sec09} are shown, all with $n=78$. The ionization energy and the energy of $5p~^2P_{\frac{3}{2}}^{\text{o}}$~(or $6p~^2P_{\frac{3}{2}}^{\text{o}}$) of $^{87}$Rb~($^{133}$Cs) are from Ref.~\cite{Sansonetti2006}~(or Ref.~\cite{Sansonetti2009}), and its quantum defects for the Rydberg states are from Ref.~\cite{Li2003}~(or Ref.~\cite{CsEnergyPhysRevA.35.4650}) for calculating the wavelengths of lasers. Masses of atoms are from Ref.~\cite{Sansonetti2005}. Quantum defects for barium are from Ref.~\cite{Neukammer1988}. \label{table-0}  }
   \begin{tablenotes}
        \item[a] Here the refractivity of air is not considered, while the wavelengths over 200~nm shown in Refs.~\cite{Sansonetti2006,Sansonetti2009} are those in air.
    \end{tablenotes}
     \end{threeparttable}
\end{table*}

An effective approach to enhancing Rydberg coherence can be with barium atoms. Barium has a metastable low-lying d-orbital $^1D_2$ clock state from which the $6sng~^1G_4$ Rydberg states can be excited with a tiny $\Bbbk\approx0.14~\mu m^{-1}$ via two counter-propagating lasers of 649~nm and 658~nm. Compared to the smallest possible $\Bbbk$ with rubidium or cesium and also due to the relatively large mass of barium, the intrinsic $T_{2,\text{\tiny D}}$ by using barium can be enhanced by about 40 times with similar motional temperature $T_{\text{eff}}$, which effectively removes the motional dephasing of Rydberg states. Further, compared to usually employed alkali-metal atoms as shown in Table~\ref{table-0}, the $6sng~^1G_4$ barium Rydberg state possesses long radiative lifetimes. This brings hope to improve Rydberg-mediated gate fidelity based on atomic qubits~\cite{PhysRevLett.85.2208,Lukin2001,Saffman2010,Saffman2016,Saffman2019nsr,Shi2021qst}, makes it possible to realize all optical quantum information based on Rydberg polaritons\cite{Paredes-Barato2014,Shao2024review}, and can improve the controllability in Rydberg-mediated quantum many-body physics~\cite{Saffman2010,Browaeys2020,Adams2020,Morgado2021,Wu2021,Kaufman2021}.

Beside of long Rydberg $T_{2,\text{\tiny D}}$ and $T_1$, there are two more strengths in the $6sng~^1G_4$ barium Rydberg state and its excitation from a low-lying $^1D_2$ clock state. (i) Atoms in the $6sng~^1G_4$ state will experience a strong dipole-dipole interaction due to very small F\"{o}rster defect with the nearby $6snf^1F_3^{\text{o}}$ Rydberg state~\cite{Neukammer1988}. (ii) For the low-lying $^1D_2$ clock state from which we excite the $6sng~^1G_4$ Rydberg state, it has a strong electric quadrupole allowed transition to the ground state, enabling fast single-qubit operations if we encode qubit with the ground and the low-lying $^1D_2$ clock states.

One thing we need to consider is the relatively short radiative lifetime of the low-lying $^1D_2$ clock state, in the range of $[242,296]$~ms~\cite{lifetime1D2}. Thankfully, there is a nearby low-lying $^3D_3$ clock state which has a lifetime of several tens of second, and can be excited from the low-lying $^1D_2$ clock state with a strong two-photon Raman transition. Using the $^3D_3$ clock state as a shelving state during the state detection and idle times can avoid extra radiative decay from the low-lying $^1D_2$ clock state.

The remainder of this paper is organized as follows. In Sec.~\ref{sec02}, we introduce the basics about barium qubits, including definition of the qubit state and the shelving, cooling, and single-qubit control. In Sec.~\ref{sec08}, we study the excitation of $6sng~^1G_4$ barium Rydberg states which has long $T_{2,\text{\tiny D}}$ and long radiative lifetimes. In Sec.~\ref{sec10}, we show that $6sng~^1G_4$ barium Rydberg states exhibit strong dipole-dipole interactions. In Sec.~\ref{polariton}, the prospect on barium-based quantum optics is discussed, and Sec.~\ref{sec12} discusses the possibility to simultaneously trap the two qubit states and the $6sng~^1G_4$ Rydberg state. We briefly summarize in Sec.~\ref{sec13}. 

\begin{figure}
\includegraphics[width=3.0in]
{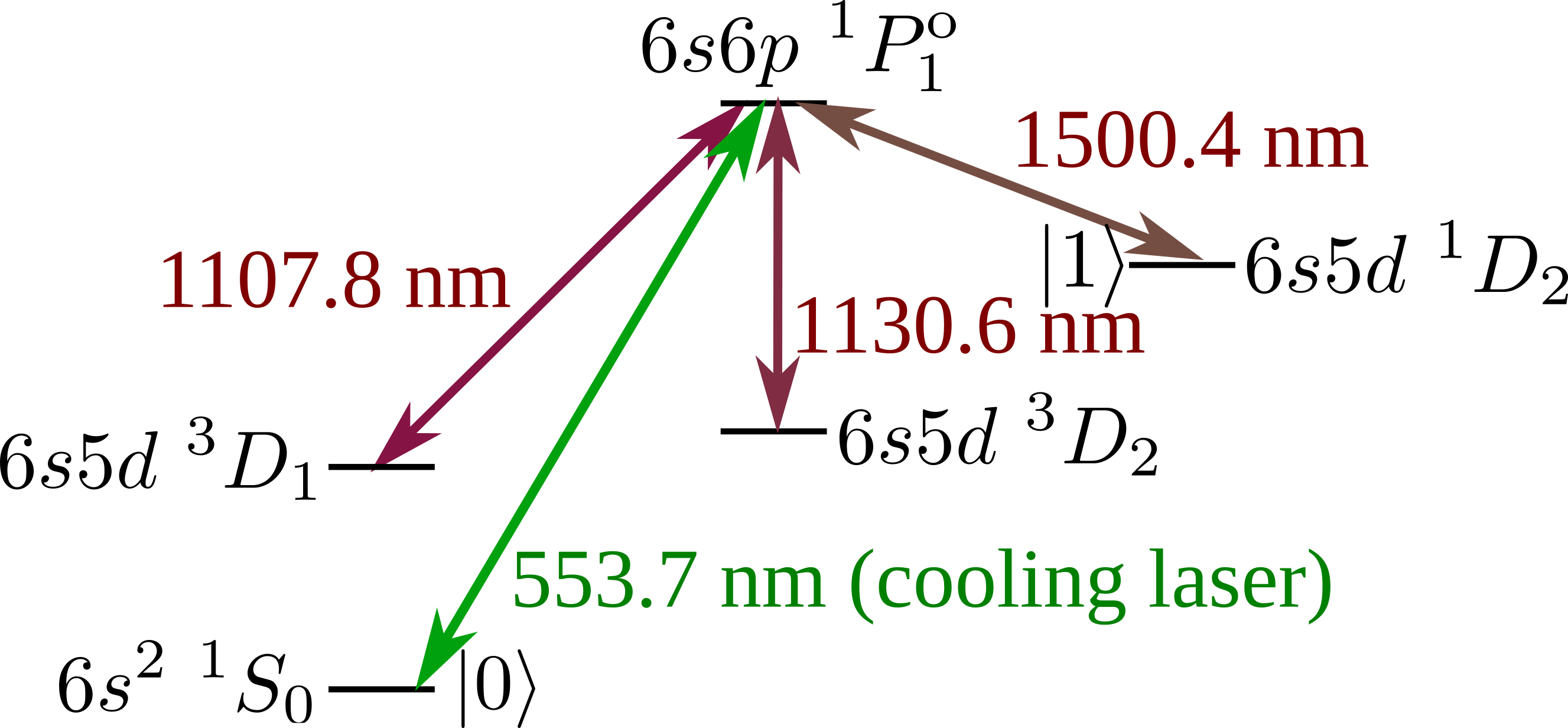}
\caption{Level scheme of the cooling and magneto-optical trapping of barium atoms used in Ref.~\cite{BarimTrap2009}. The level spacing is not to scale. \label{figure-trap} }
\end{figure}
\section{Barium qubit and its quantum control}\label{sec02}
Barium has seven naturally occurring isotopes, among which $^{138}$Ba has the largest abundance~\cite{Sansonetti2005}. In this work, we consider atomic qubits defined by $^{138}$Ba which has no hyperfine structure. The energy levels and lifetimes of several low states of barium in Fig.~\ref{figure-trap} are well understood~\cite{Klose2002,Curry2004,Dzuba2006,Dzuba2007,Shen2011,Yu2011,Dzuba2014,Wang2017,Qian2019}, based on which we propose to define the qubit by the ground state $6s^2~^1S_0$ and the metastable $6s5d~^1D_2$ state,
\begin{eqnarray}
\lvert 0\rangle&\equiv& \lvert 6s^2~^1S_0\rangle,\nonumber\\
\lvert 1\rangle&\equiv& \lvert 6s5d~^1D_2,m\rangle,\label{BaQubitState}
\end{eqnarray}
where the magnetic quantum number $m$ can be any one of $0,~\pm1$, and $\pm2$ because a fast electric quadrupole~(E2) transition between $|0\rangle$ and $6s5d~^1D_2$ of any $|m|\leq2$ is realizable with moderate power of a 877~nm laser as shown in the Supplemental Material~(SM)~\cite{supple}. To our knowledge, there is no experimentally measured radiative lifetime $t_{\text{radiat}}$ for $6s5d~^1D_2$, but according to a systematic theoretical study~\cite{lifetime1D2} as discussed in SM~\cite{supple}, $t_{\text{radiat}}$ is in the range of $[242,296]$~ms. Beside of $6s5d~^1D_2$, the $6s5d$ manifold of barium has several other states possessing relatively long lifetimes~\cite{supple}, but we haven't found any small wavevector for a Rydberg excitation if we consider qubit definition with them.

The radiative lifetime $t_{\text{radiat}}\in[242,296]$~ms of the qubit state $\lvert1\rangle$ can be long compared to the inhomogeneous coherence time of optically trapped optical, hyperfine, or fine-structure neutral atom qubits, which is usually on the order of millisecond~\cite{Graham2019,Picken2018,manetsch2024,SrFineQubit202401,SrFineQubit202402}. However, this $t_{\text{radiat}}$ is short for algorithms based on dynamical decoupling~\cite{Bluvstein2022,Bluvstein2023,Lukin2025topo} which can yeild a homogeneous coherence time on the order of second~\cite{Picken2018,Singh2023,Graham2022,Bluvstein2022,Wang2016,manetsch2024}. Without dynamical decoupling, however, Ref.~\cite{Graham2022} reported functional quantum circuits with running times up to 1.1~ms when the inhomogeneous coherence time of the qubit is less than $8$~ms.

\begin{figure}
\includegraphics[width=3.4in]
{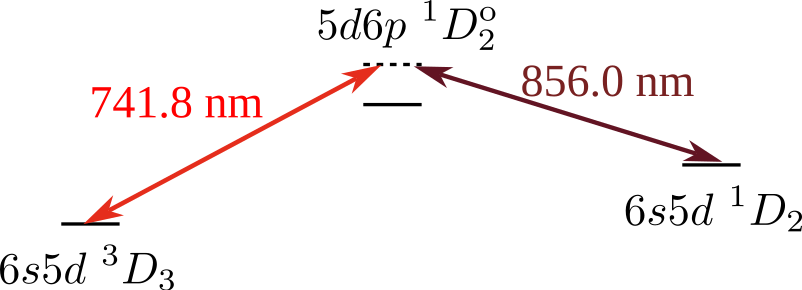}
\caption{ Level scheme of population transfer between $\lvert1\rangle$ and $6s5d~^3D_{3}$ in Eq.~(\ref{shelving01}). During idle time in a quantum circuit, e.g., between the execution of two quantum gates, a fast population transfer can be used so that the population in $\lvert1\rangle$ can be shelved on the long-lived $6s5d~^3D_{3}$ state. After the idle time, the reverse population transfer back to $\lvert1\rangle$ can be similarly realized. \label{figure-shelving} }
\end{figure}

To avoid decay error in $\lvert1\rangle$ during idle times in the execution of a quantum computing task, one can shelve the state $\lvert1\rangle$ in a long-lived state like, e.g., $6s5d~^3D_{3}$, which can have a longitudinal lifetime over 100~s~\cite{supple}. One can also shelve the state $\lvert1\rangle$ in $6s5d~^3D_{3}$ during the readout by measuring the population in $\lvert0\rangle$. By this we can avoid extra decay error from $\lvert1\rangle$. Figure~\ref{figure-shelving} shows the scheme for the coherent population transfer between $\lvert1\rangle$ and $6s5d~^3D_{3}$ via the intermediate state $5d6p~^1D_2^{\text{o}}$~\cite{Niggli1987}
\begin{eqnarray}
 6s5d~^3D_{3} \xleftrightarrow[ ]{741.8~\text{nm} } 5d6p~^1D_2^{\text{o}}\xleftrightarrow[ ]{856.0~\text{nm} } 6s5d~^1D_2.\label{shelving01}
\end{eqnarray}
As shown in SM~\cite{supple}, with powers around 30~mW for the lasers of Eq.~(\ref{shelving01}), a population transfer between $\lvert1\rangle$ and $6s5d~^3D_{3}$ is possible with an infidelity of order $10^{-6}$.

The trapping and cooling of barium was realized in Ref.~\cite{BarimTrap2009} based on the strong $6s^2~^1S_0-6s6p~^1P_1^{\text{o}}$ transition as illustrated in Fig.~\ref{figure-trap}. The cooling cycle of Ref.~\cite{BarimTrap2009} has a Doppler limit of 0.44~mK~\cite{Dammalapati-thesis}. To further cool the atoms, Ref.~\cite{Dammalapati2012} proposed to use the intercombination transition $6s^2~^1S_0-6s6p~^3P_1^{\text{o}}$, which has a Doppler cooling limit 2.9~$\mu$K. However, the state $6s6p~^3P_1^{\text{o}}$ can decay back to $6s5d~^3D_{1,2}$ and $6s5d~^1D_{2}$, so repumping lasers are necessary. Alternatively, SM~\cite{supple} shows that it is possible to realize gray molasses cooling, which can also significantly reduce the temperature of atoms following the first-stage cooling.

The atoms trapped in the magneto-optical trap can be loaded into optical tweezers stochastically. We find that there are several magic wavelengths for trapping both the ground and the $6s5d~^1D_2$ states with a $\pi$ polarized laser field~\cite{supple}, with dynamic polarizability of similar order to those of ytterbium~\cite{Yamamoto2016,Guo2010}. One limitation to the qubit coherence is the Raman scattering of the tweezer light~\cite{Saffman2005,Ma2023,PhysRevX.13.041035}. There are two magic wavelengths around 885~nm and 1634~nm, both of which have a small scattering rate below $2\pi\times10$~mHz with a 1~mK trap depth~\cite{supple}. After loading the atoms into tweezers, another set of optical tweezers can be used to rearrange the array so that it becomes defect-free. With the mature techniques for building defect-free neutral-atom arrays~\cite{Endres2016,Kim2016,Lin2024,Adams2020,Chiu2025}, it is possible to form tweezer array of barium qubits based on the cooling scheme of~\cite{BarimTrap2009,Dammalapati2012} or the gray molasses cooling~\cite{supple}.

Single-qubit gates can be realized by a fast transition between $6s^2~^1S_0$ and $6s5d~^1D_2$. A direct transition between the qubit states is not electric dipole allowed, but according to the selection rules for electric quadrupole transitions in a hydrogen-like atom,
it is possible to realize population transfer between the ground state and $6s5d~^1D_2$ directly by a 877.6~nm laser via an E2 transition. As detailed in SM~\cite{supple}, this E2 transition has an electric quadrupole matrix element over 17$ea_0^2$, and Rabi frequencies over $2\pi\times$20~MHz are realizable with laser power below 40~mW if the laser waist is around $10~\mu$m. Based on this E2 transition, single-qubit gates are readily realizable with high accuracy~\cite{supple}.

\section{Rydberg excitation with long Doppler dephasing time and Rydberg lifetime}\label{sec08}
\subsection{Long $T_{2,\text{\tiny D}}$}
There are several ways for producing Rydberg atoms, like charge exchange, electron impact, photoexcitation, and even a combination of collisional and optical excitation~\cite{Gallagh2005}. Here, we consider photoexcitation which can provide coherent manipulation of the atomic qubits in the array. Rydberg states of barium atoms were well examined in experiments~\cite{PhysRevA.18.2173,Armstrong79,PhysRevA.21.588,PhysRevA.26.379,Neukammer1982,Neukammer198202,PhysRevA.26.2611,PhysRevA.27.2995,PhysRevA.28.850,PhysRevA.28.1901,PhysRevA.29.2989,PhysRevLett.59.2947,Neukammer1988,PhysRevA.58.3058,PhysRevA.64.033409,PhysRevA.76.052505,PhysRevA.82.042507,PhysRevA.85.032508,PhysRevA.85.052517,PhysRevA.89.062503,Shen2011}. Here, we consider two counter-propagating lasers to excite the following transition
\begin{eqnarray}
 6s5d~^1D_2 \xleftrightarrow[\Omega_{\text{d-f}}]{648.5~\text{nm} } 5d6p~^1F_3^{\text{o}} \xleftrightarrow[\Omega_{\text{f-g}}]{657.9~\text{nm} } 6s78g~^1G_4,\label{Ba-Rydberg}
\end{eqnarray}
with a large one-photon detuning at $5d6p~^1F_3^{\text{o}}$. Here, the principal quantum number $n=78$ is shown for it is the highest $^1G_4$ barium Rydberg state with the energy measured~\cite{Neukammer1988}. Note that the wavelength of the left transition above differs a little from, e.g., that in Ref.~\cite{Wang2017}, possibly because some data is for light in vacuum while some other is for light in air.

By sending the two laser lights of Eq.~(\ref{Ba-Rydberg}) in a counter-propagating configuration, the two-photon wavevector is ${\Bbbk}\approx0.138~\mu$m$^{-1}$, which is 36 times smaller than the smallest possible wavevector of a two-photon Rydberg excitation with rubidium or cesium, i.e., $\Bbbk\approx5.0~\mu m^{-1}$ with two counter-propagating lasers of 852~nm and 509~nm for cesium~\cite{Shi2020prapplied}. Moreover, the relatively heavy mass of barium adds another factor for decreasing the Rydberg dephasing. To put it in perspective, we note that if the laser waist is large compared to the distance for the atom to move during the excitation~\cite{Jenkins2012}, the coherence time of the laser-excited Rydberg state is given by~\cite{Wilk2010,Saffman2011,Jenkins2012,PhysRevA.101.013421}$\frac{1}{\Bbbk}\sqrt{\chi\frac{m}{k_{\text{\tiny B}}T_{\text{eff}}}}$,
with $m$ the mass of the atom and $k_{\text{\tiny B}}$ the Boltzmann’s constant. Here, $\chi=2$ when Doppler dephasing dominates the motional dephasing in atom entanglement of an atom array~\cite{Wilk2010,Saffman2011,Graham2019}, and $\chi=1$ when both Doppler dephasing and photon recoil are significant in Rydberg EIT-based quantum nonlinear optics in atom gas~\cite{Jenkins2012,PhysRevA.101.013421}. Photon recoil also limits gate fidelity in atom array~\cite{Robicheaux2021}, so we can consider $\chi=1$ to estimate an upper bound of the Rydberg coherence time, i.e.,
\begin{eqnarray}
 T_{2,\text{\tiny D}} &= & \frac{1}{\Bbbk}\sqrt{ \frac{m}{k_{\text{\tiny B}}T_{\text{eff}}}},\label{eq-T2}
\end{eqnarray}
which shows that the wavevector has a profound influence on $T_{2,\text{\tiny D}}$ though the mass of the atom does matter. In Table~\ref{table-0} we show the smallest possible $\Bbbk$ for two-photon Rydberg excitation of ground-state cesium or rubidium. For comparison, Table~\ref{table-0} shows the corresponding $\Bbbk$ in Eq.~(\ref{Ba-Rydberg}). According to Eq.~(\ref{eq-T2}), the fourth column of Table~\ref{table-0} shows that at the same atomic temperature, the motion-induced Doppler dephasing of $6sng~^1G_4$ barium Rydberg states is about 47~(37) times slower than the slowest possible ground-Rydberg dephasing rate of rubidium~(cesium).

\subsection{Long radiative lifetime}\label{sec09}
We didn't find published data about the radiative lifetime of Barium Rydberg state $6sng~^1G_4$. To analyze it, we note that the actual level configuration or state mixing matters. In detail, the $6snl$-orbital Rydberg states of barium can be perturbed by nearby $5dml$~(such as $5d7d,5d8p,\cdots$) states. However, unlike the three triplet $6sng~^3G_{3,4,5}$ Rydberg states which are perturbed by nearby $5d7d$ states, the singlet $6sng~^1G_4$ Rydberg state is not perturbed~\cite{PhysRevA.28.850}. This offers relatively simple analysis about its property such as the radiative lifetime. So, the radiative coupling with the $^1F_3$ and $^1H_5^{\text{o}}$ states contributes to the $T_1$ relaxation of a $6sng~^1G_4$ Rydberg state.

Another factor that matters in the analysis of Rydberg state lifetime is the energy of states. For Rydberg states of the form $6sng~^1G_4$, $6snf~^1F_3$, and $6snh~^1H_5^{\text{o}}$, the quantum defect was measured experimentally~\cite{Armstrong79,Neukammer1988,Neukammer1988}. Multichannel quantum-defect analysis also provided quantum defects~\cite{PhysRevA.28.850}. As an estimate, we use the results of Ref.~\cite{Neukammer1988}. Note that unlike the three triplet $6sng~^3G_{3,4,5}$ Rydberg states, the singlet $6sng~^1G_4$ Rydberg state is not perturbed by nearby $5d7d$ states~\cite{PhysRevA.28.850}, therefore we can use quantum defects to calculate the energy of the actual $6sng~^1G_4$ Rydberg state.

\begin{figure}
\includegraphics[width=2.4in]
{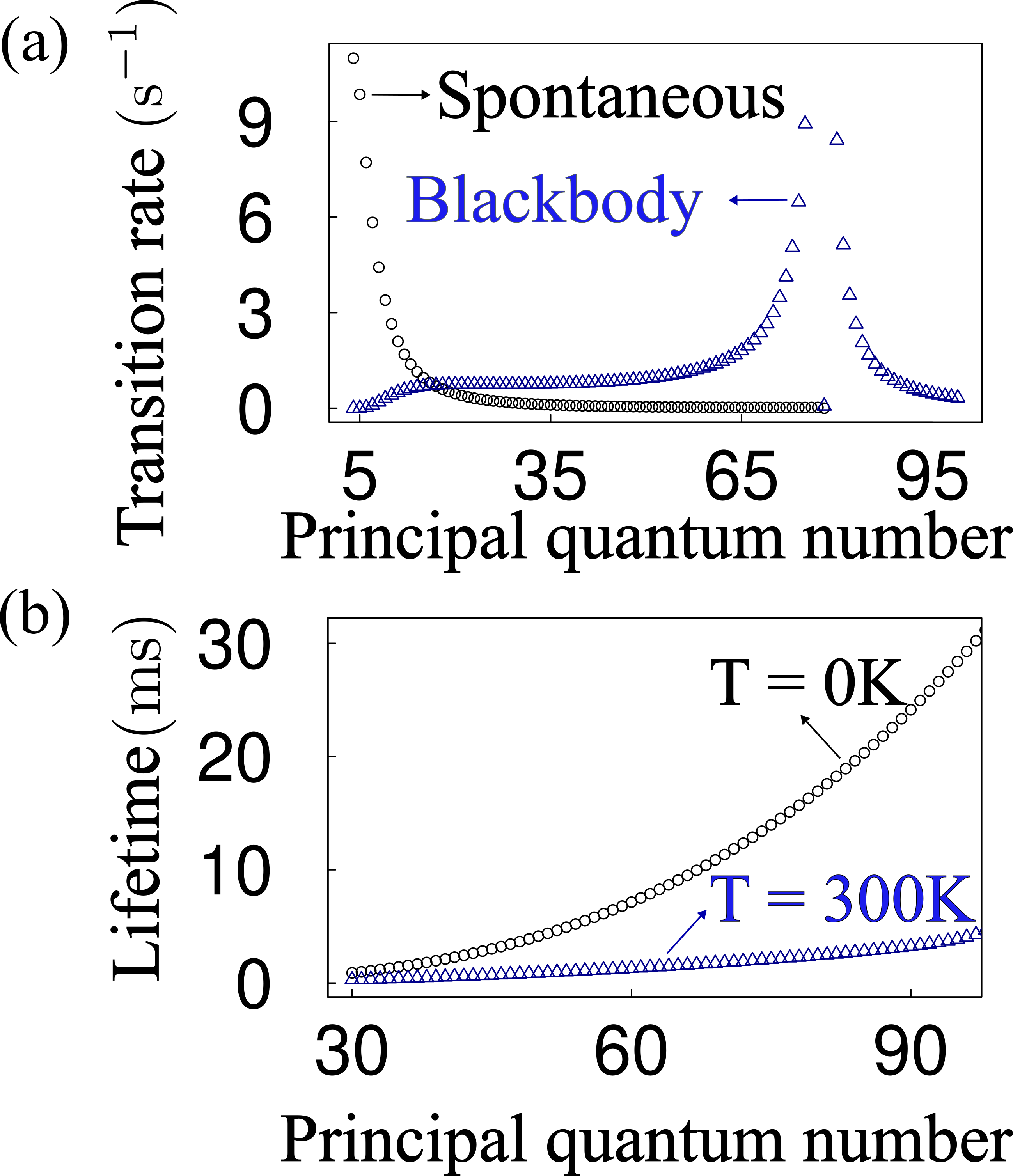}
\caption{(a) Round and triangle symbols denote spontaneous emission rate and black-body-induced radiation rate from $6s78g~^1G_4$ to $6snf~^1F_3^{\text{o}}$ and $6snh~^1H_5^{\text{o}}$, respectively. The round symbols disappear beyond principal quantum number 78 for spontaneous emission only takes atom to a lower state. (b) Round and triangle symbols denote $1/\Gamma_0$ and $1/(\Gamma_0  + \Gamma_{\text{\tiny BBR} })$, respectively, where blackbody transition to states up to $n=100$ is included.   \label{figure-G2F} }
\end{figure}

As detailed in SM~\cite{supple}, we use the semiclassical expression for the radial matrix elements~\cite{Kaulakys1995} in the calculation of the radiative lifetime. The semiclassical approach has been tested in the study of high Rydberg states~\cite{Walker2008}. Figure~\ref{figure-G2F}(a) shows the transition rate from $6smg~^1G_4$ with $m=78$ to a $6snf~^1F_3^{\text{o}}$ state, where the round and triangle symbols denote contributions from spontaneous emission and blackbody radiation, respectively. By summing over transitions to both $6snf~^1F_3^{\text{o}}$ and $6snh~^1H_5^{\text{o}}$, we have evaluated the radiation lifetimes of the state $6sng~^1G_4$ at both zero and 300 K of principal quantum number $n$ over 30, shown in Fig.~\ref{figure-G2F}(b). The lifetimes of $6sng~^1G_4$ barium Rydberg states with $n=60$ are about 7.1~ms and 1.3~ms at 0 and 300~K, respectively, which is long compared to those of low-angular-momentum Rydberg states of alkali atoms~\cite{Beterov2009}.

\subsection{Prospects on Rydberg entangling gates } \label{secVC}
Here, we compare entangling gate fidelities between barium-based neutral-atom quantum computing with those based on akali-metal atoms~\cite{Evered2023}, ytterbium~\cite{Ma2023}, and strontium~\cite{tsai2024fid}. 

We consider $n=53$ for the comparison with Ref.~\cite{Evered2023} where the $53S_{1/2}$ Rydberg state was employed. As analyzed in SM~\cite{supple}, by assuming laser waist around 10~$\mu$m, one can have $\Omega_{\text{f-g}}/2\pi = 200$~MHz and $\Omega_{\text{d-f}}/2\pi = 30$~GHz with laser powers about 2.0~W and 0.2~W for the two transitions in Eq.~(\ref{Ba-Rydberg}), respectively. To avoid scattering at $\lvert5d6p~^1F_3^{\text{o}}\rangle$, the two fields for inducing the two transitions in Eq.~(\ref{Ba-Rydberg}) shall have a large detuning $\Delta_{\text{f}}$ at $\lvert5d6p~^1F_3^{\text{o}}\rangle$. If a similar gate protocol as the time-optimal gate protocol of Ref.~\cite{Evered2023} is used with barium, the gate error due to the scattering at $\lvert5d6p~^1F_3^{\text{o}}\rangle$ is around $3\times10^{-4}$ with $|\Delta_{\text{f}}/\Omega_{\text{f-g}}|=100$~\cite{supple}. This scattering error is smaller than that of Ref.~\cite{Evered2023} shown in Extended Data Table 1 therein. But the two major error sources in the gate of Ref.~\cite{Evered2023} are Rydberg-state decay and dephasing. Table~\ref{table-0} shows that the Doppler dephasing is much suppressed with barium, and from Sec.~\ref{sec09} we find that the lifetime is about 990~$\mu$s for the $6s53g~^1G_4$ state of barium, which is about ten times longer than that of a Rydberg state of the same principal quantum number in~\cite{Evered2023}. This means that barium qubit can help to realize high-fidelity CZ gate.

In the extended data Table~2 of Ref.~\cite{Ma2023}, it shows that their CZ gate has an error of $0.02$, half of which is from the Rydberg $T_1$ and the Rydberg Doppler shift. In Ref.~\cite{tsai2024fid}, the analysis on the gate fidelity in its Fig.~1(c) shows that Rydberg-state decay contributes about an error of 0.001, and the other major error is from the frequency noise of the UV laser. The lifetime is 65~$\mu$s~(n=59) and $\lesssim78~\mu$s~(n=61) for the Rydberg state used in Ref.~\cite{Ma2023} and Ref.~\cite{tsai2024fid}, respectively, while by the method of Sec.~\ref{sec09}, the lifetime of barium $^1G_4$ state is about 1300~$\mu$s at similar n. On the other hand, the frequency noise of optical laser can be well suppressed~\cite{Evered2023}. These show that it is possible to realize high-fidelity CZ gate with barium atoms.

\subsection{Comparison with existing Doppler-robust schemes}
It is useful to compare the current theory with typical Doppler-robust CZ gate schemes in, e.g., Refs.~\cite{Fromonteil2023,Jandura2023}.

Both Ref.~\cite{Fromonteil2023} and Ref.~\cite{Jandura2023} proposed to reverse the Doppler shift either by inverting the direction                                                                                                                                                                    of the Rydberg laser or by inverting the direction of the atomic velocity along the Rydberg laser. The latter approach is based on that the motion of the atom in the harmonic optical tweezer potential is periodic, so the velocity of each atom is reversed after each half of the trap period. To mitigate several disadvantages associated with this approach of inverting the velocity of the atomic motion, Ref.~\cite{Jandura2023} further proposed to modulate the trapping potential sinusoidally in time. For the approach of inverting the direction of the Rydberg laser, the relative path lengths of the two different Rydberg lasers require experimental calibration. Both Ref.~\cite{Fromonteil2023} and Ref.~\cite{Jandura2023} designed novel Doppler robust gates by executing two Rydberg pulses, where the Doppler shift in the first pulse is opposite to that in the second pulse. A compromise is that the gate duration would be a little longer compared to their gate schemes without considering suppressing Doppler shift. In comparison, the theory in this paper does not depend on reversing the Doppler shift, so that using the fast gate schemes in, e.g., Ref.~\cite{Evered2023}, would be enough to suppress the Doppler shift for our case.

The suppression of the Doppler dephasing by the theory in this work depends on the cooling of atoms, while the Doppler robust gates in Refs.~\cite{Fromonteil2023,Jandura2023} appear not so dependent on cooling of atoms. For example, Fig.~5(b) of Ref.~\cite{Jandura2023} shows that their Doppler robust gate nearly has a constant infidelity $10^{-4}$ for motional temperature $T_{\text{eff}}$ from 0~$\mu$K to $45~\mu$K. For the theory in this paper, Eq.~(\ref{eq-T2}) shows that $T_{2,\text{\tiny D}}~\approx 933~\mu$s~$\sqrt{\mu \text{K}/ T_{\text{eff}}}$. This means that $T_{2,\text{\tiny D}}$ would be up to $100~\mu$s even with $T_{\text{eff}}=80~\mu$K. If we use a gate protocol as in Ref.~\cite{Evered2023}, then the gate error due to the Rydberg dephasing can be analyzed as Ref.~\cite{Evered2023}, where a Rydberg dephasing time $3~\mu$s contributed an error $0.134\%$, which means that a Rydberg dephasing time $100~\mu$s would incur an error of about $4\times10^{-5}$ if we use the barium qubits in the CZ gate of Ref.~\cite{Evered2023}. This estimate depends on large Rydberg Rabi frequencies up to $2\pi\times4.6$~MHz as in Ref.~\cite{Evered2023}, where the possibility is shown in Sec.~\ref{secVC} with more details in SM~\cite{supple}. The error $4\times10^{-5}$ is negligible compared to other error sources according to recent experiments~\cite{Evered2023,Ma2023,tsai2024fid}. This means that the theory in this work requires cooling the barium atoms to $80~\mu$K or below so as to provide a practical improvement.

The theory in this paper is advantageous in suppressing the Rydberg-state decay. The Rydberg-state decay can be captured by the Rydberg superposition time $t_{\text{\tiny Ryd}}$ averaged over the different input states of the gate. The fully Doppler robust gate~(Protocol III therein) in Ref.~\cite{Fromonteil2023} has a Rydberg superposition time $t_{\text{\tiny Ryd}}\approx17/\Omega_{\text{\tiny Ryd}}$. In Ref.~\cite{Jandura2023}, the Doppler robust gate has a Rydberg superposition time $t_{\text{\tiny Ryd}}\approx5.6/\Omega_{\text{\tiny Ryd}}$, and the erasure-conversion based conditionally amplitude- and Doppler-robust gate has a Rydberg superposition time $t_{\text{\tiny Ryd}}\approx6.6/\Omega_{\text{\tiny Ryd}}$. For the theory in this work, using the gate protocol of Ref.~\cite{Evered2023} is enough, which has $t_{\text{\tiny Ryd}}\approx3.0/\Omega_{\text{\tiny Ryd}}$. Moreover,
the Rydberg lifetime $\tau$ considered in this paper is about 10 to 20 times of those in the alkali-metal, ytterbium, or strontium setup as discussed in Sec.~\ref{secVC}, so that the gate infidelity $t_{\text{\tiny Ryd}}/\tau$ due to Rydberg-state decay with barium qubits would be orders of magnitude smaller. Of course, if the decay of Rydberg state can take the atom to metastable states other than the qubit states, like in Ref.~\cite{Ma2023} where the qubit was encoded in the clock state of $^{171}$Yb, the part of the decay to metastable states other than the qubit states can be corrected in the context of mid-circuit erasure conversion~\cite{Jandura2023}. This in principle means that the decay of the barium Rydberg states to the several metastable $6s5d$ states in Fig.~\ref{figure-trap} can be corrected. However, this may not be necessary due to the extremely long lifetime of barium $6s53g~^1G_4$.


Finally, it is unclear whether the theory in Refs.~\cite{Fromonteil2023,Jandura2023} can suppress motional dephasing of Rydberg polaritons. The theory in this article is based on small wavevector of the Rydberg excitation, and thus is useful for suppressing the fast motional dephasing of Rydberg polaritons which is of practical importance in Rydberg-mediated quantum nonlinear optics~\cite{Dudin2012,Peyronel2012,Firstenberg2013,Baur2014,Tiarks2014,Tiarks2018,PhysRevLett.127.063604}.

\section{Strong Rydberg interaction }\label{sec10}
A special character of the Rydberg interaction of the $6sng~^1G_4$ barium state is the strong F\"{o}ster resonance due to the near degeneracy of $6snf^1F_3^{\text{o}}$ and $6sng~^1G_4$~\cite{Neukammer1988}. This will result in strong interaction in two atoms if both of them are in the $6sng~^1G_4$ state. There are two cases, (i) when the two atoms are near enough so that the dipole-dipole interaction is comparable or larger than the energy defect~\cite{Walker2008}, and (ii) when the two atoms are far so that a dispersive energy shift occurs to the two-atom state.

\subsection{Near-resonant interaction }
For barium, the Rydberg state $6sng~^1G_4$ is quite near to $6snf^1F_3^{\text{o}}$, and a little near to $6snh~^1H_5^{\text{o}}$, as experimentally measured in Ref.~\cite{Neukammer1988}. For example, the gap between $6snf^1F_3^{\text{o}}$ and $6sng~^1G_4$ is $0.01$cm$^{-1}$ at $n=57$, or 300~MHz, and the gap shrinks when $n$ increases. Meanwhile, Ref.~\cite{Neukammer1988} showed that the gap between $6snh~^1H_5^{\text{o}}$ and $6sng~^1G_4$ is relatively larger, and, e.g., is 1.4 and 0.6~GHz at $n=57$ and 78, respectively. This means that the interaction between two atoms in the state $\lvert 6sng~^1G_4\rangle \otimes \lvert 6sng~^1G_4\rangle$ can be strong due to the small energy defect $\delta_{\text{fg}}$~(i.e., energy difference) between $\lvert 6sng~^1G_4\rangle \otimes \lvert 6sng~^1G_4\rangle$ and $\lvert 6snf~^1F_3^{\text{o}}\rangle \otimes \lvert 6snf~^1F_3^{\text{o}}\rangle$, where $\delta_{\text{fg}}$ was defined following Ref.~\cite{Walker2008}.

\begin{figure}
\includegraphics[width=3.4in]
{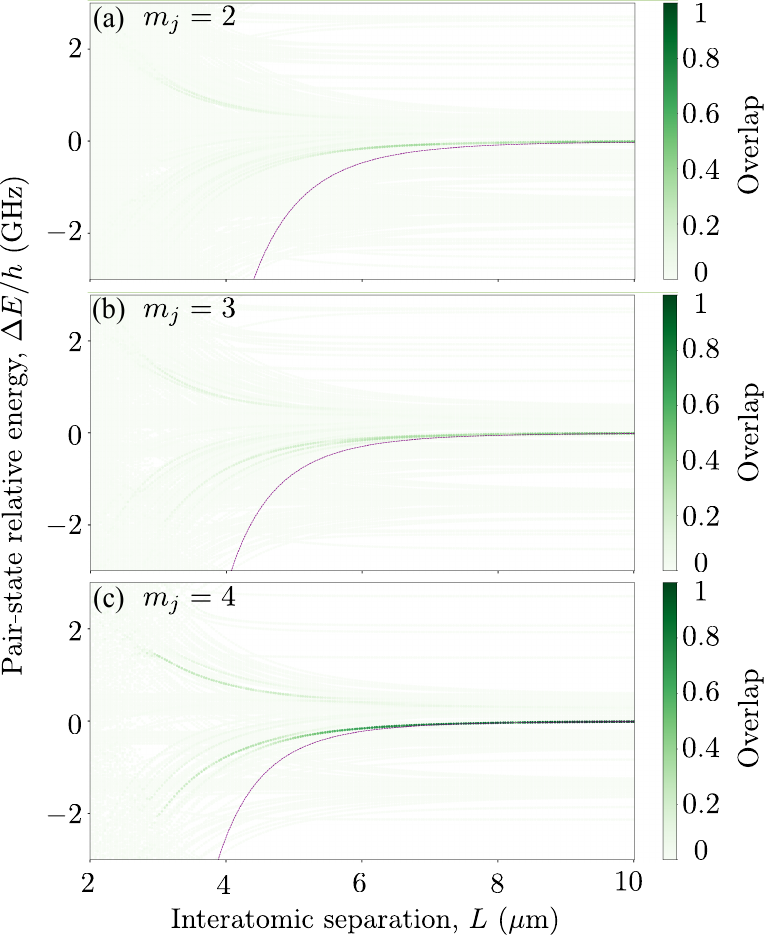}
\caption{Interaction strengths of the dipole-dipole interaction between two atoms in the state $|ngng\rangle\equiv\lvert 6sng~^1G_4, m_j\rangle \otimes \lvert 6sng~^1G_4,m_j\rangle$ with $m_j=2,~3$, and $4$ in (a), (b), and (c), respectively. Here $n=78$ is taken as an example because it is the highest $^1G_4$ barium state with the energy measured~\cite{Neukammer1988}. Here $\theta= 0$, the calculation was via ARPACK~\cite{arpack}, and states within $25$~GHz from $|ngng\rangle$, $|\Delta n|\leq 4$, and $s,p,\cdots,i,k$-orbitals are included. The purple dashed curve shows the asymptotic van der Waals scaling $C_6/L^6$ with perturbatively calculated $C_6$.     \label{figure-C3gf0} }
\end{figure}

The dipole-dipole interaction can couple a huge number of pair states to any initial pair states of two atoms. For two atoms in $\lvert 6sng~^1G_4, m_j\rangle \otimes \lvert 6sng~^1G_4,m_j\rangle$ which is abbreviated as $|ngng\rangle$, the dipole-dipole interaction will couple it with $|6sn_1f,6sn_2f\rangle,|6sn_3h,6sn_4h\rangle,$ and $|6sn_5f,6sn_6h\rangle$ states, where $n_{j}$ with $j=1-6$ represents a principal quantum number around $n$, and these latter states will be coupled to other states. As shown in 
Table~SM-I of SM~\cite{supple}, the first four eigenstates of the dipole-dipole coupling descending in overlap with $|ngng\rangle$ can have d- and i-orbital states, where we show examples with $|m_j|=4$ for such states are less sensitive to noise of electric field~\cite{supple}. Thus, it is necessary to include enough states near to $|ngng\rangle$ when studying the dipole-dipole interaction.

With $\theta=0$ and by truncating states $25$~GHz~\cite{Robertson2021}
away from $|ngng\rangle$, allowing change of principal quantum number up to 4, and including angular states $s,p,\cdots,i,k$, there are $\mathcal{N}$=6218, 4237, and 2400 pair states coupled when $|m_j|=2, 3$, and 4, respectively. Using ARPACK~\cite{arpack}, one can diagonalize the $\mathcal{N}\times\mathcal{N}$ matrix of the interaction Hamiltonian, resulting in $\mathcal{N}$ eigenstates $\lvert \mu_{\beta}\rangle$, where the eigenvalues are shown in Figs.~\ref{figure-C3gf0}(a),~\ref{figure-C3gf0}(b), and~\ref{figure-C3gf0}(c) for $m_j=2, 3$, and 4, respectively.


\begin{figure}
\includegraphics[width=3.4in]
{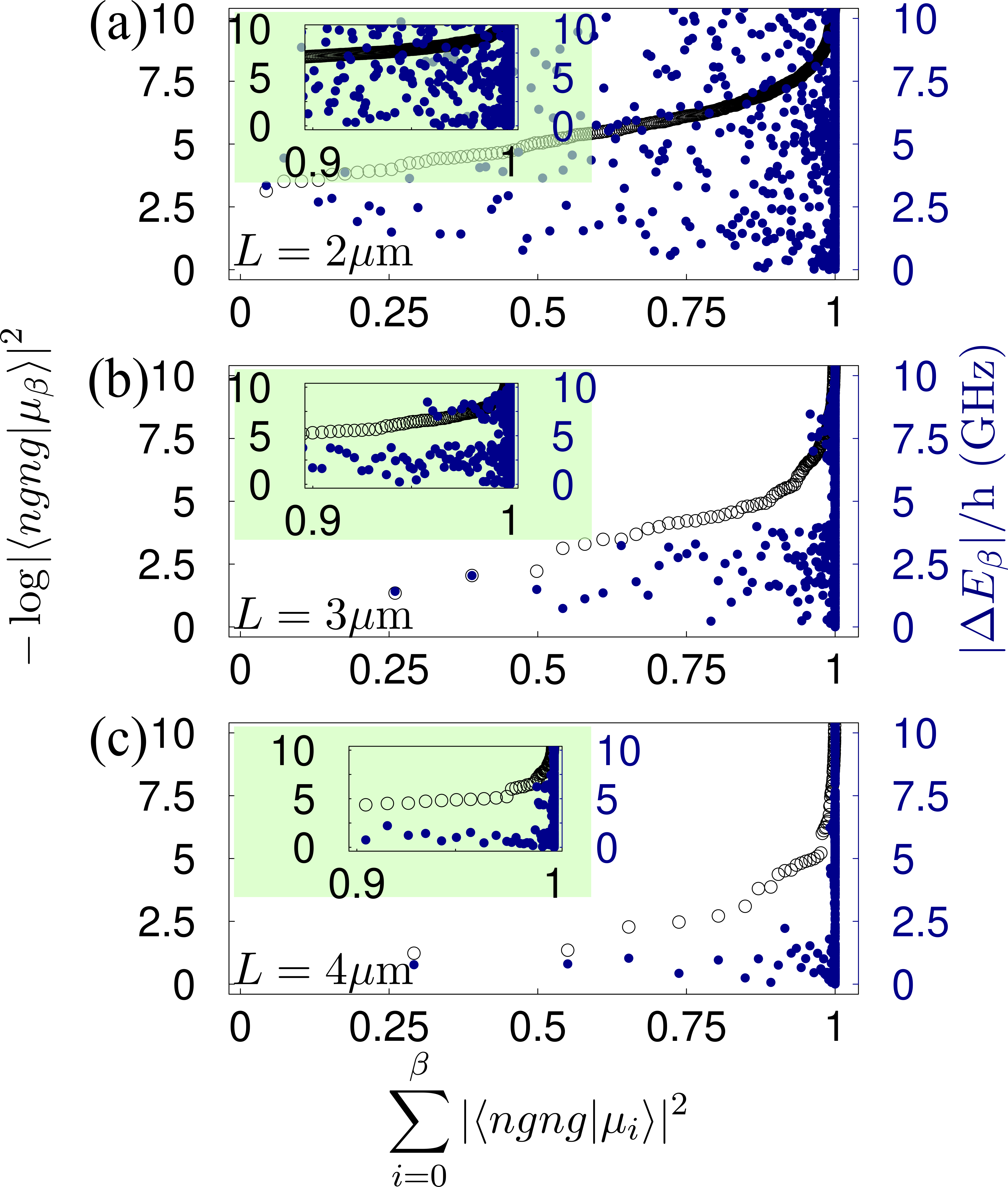}
\caption{Illustration of eigenstates of the dipole-dipole Hamiltonian for the state $|ngng\rangle$, where the empty and filled circles represent $-\log\lvert\langle ngng|\mu_{\beta}\rangle|^2$ and the pair-state energy $\Delta E$ as a function of $s_\beta=\sum_{i=0}^\beta\lvert\langle ngng|\mu_{i}\rangle|^2$. The states are ordered with descending $\lvert\langle ngng|\mu_{\beta}\rangle|^2$. The internuclear separation is $2,~3$, and 4~$\mu$m in (a,~b), and (c), respectively. The insets show the data near the region of $s_\beta=1$, which shows that $|\mu_{\beta}\rangle$ with small $\Delta E_\beta$ has small population. Here, the states have $|m_j|=4$ for they can be more useful for being less sensitive to noise of electric field~\cite{supple}.     \label{figure-C3gf} }
\end{figure}

In Rydberg-mediated quantum science~\cite{Shi2021qst}, the blockade mechanism~\cite{PhysRevLett.85.2208} requires strong energy shift to the pair state, where the blockade interaction is characterized by the eigenvalue and the overlap $\lvert\langle ngng|\mu_{\beta}\rangle|^2$ for each $\lvert \mu_{\beta}\rangle$. In Fig.~\ref{figure-C3gf0}, the color denotes the overlap of the corresponding eigenstate with $|ngng\rangle$. Ideally, if $\Delta E$ is large for all $|\mu_{\beta}\rangle$, the blockade mechanism is preserved. But if $\Delta E$ is small for a $|\mu_{\beta}\rangle$ where $\lvert\langle ngng|\mu_{\beta}\rangle|^2$ is large, then there is a blockade leakage. To examine whether there is such a concern for barium, we take states with $m_j=4$ as an example and use empty circle and filled circle to show $-\log\lvert\langle ngng|\mu_{\beta}\rangle|^2$ and the pair-state energy as a function of $s_\beta=\sum_{i=0}^\beta\lvert\langle ngng|\mu_{i}\rangle|^2$ in Fig.~\ref{figure-C3gf}. In Fig.~\ref{figure-C3gf}, one can see that the energy shift $\Delta E$ is large for most states. By sorting the numerical data descending with $\lvert\langle ngng|\mu_{\beta}\rangle|^2$, we find that for Fig.~\ref{figure-C3gf}(a) with $L=2~\mu$m, $\Delta E/h$ is larger than 1~GHz until $\Delta E/h\approx776$~MHz at $j=29$, but at such a $j$ we have a tiny overlap $\lvert\langle ngng|\mu_{\beta}\rangle|^2\approx 7.5\times10^{-3}$. Among the 2400 eigenstates, there are only four meaningful states with $\Delta E/h<100$~MHz, and these four eigenstates have $\lvert\langle ngng|\mu_{\beta}\rangle|^2=(9.6,2.7,0.92,0.27)\times10^{-4}$ and $\Delta E/h=(72,23, 7.6,67)$~MHz, respectively, where `meaningful' means that there are several eigenstates with $\Delta E/h<100$~MHz but they have $\lvert\langle ngng|\mu_{\beta}\rangle|^2<10^{-20}$. The very small $\Delta E/h=7.6$~MHz seemingly indicates a blockade leakage channel, but the corresponding eigenstate only has an overlap $9.2\times10^{-5}$ with $\lvert ngng\rangle$. To further characterize the strength of the interaction, we define an average blockade strength
\begin{eqnarray}
\overline{\Delta E} = \sum_{i=0}^\mathcal{N} \lvert\langle ngng|\mu_{i}\rangle|^2\Delta E_i,
\end{eqnarray}
and find that it is $h\times 6.1~$GHz at $L=2~\mu$m.

When $L$ increases, more solid circles in Fig.~\ref{figure-C3gf}(b,c) tend to lower down, which indicates that the interaction becomes weaker. But at $L=3~\mu$m, there are only three $|\mu_{\beta}\rangle$ with $\Delta E/h$ lower than 100~MHz, with $\lvert\langle ngng|\mu_{\beta}\rangle|^2=(4.0,0.069,0.052)\times10^{-4}$ and $\Delta E_{\beta}/h=(84,88,9.2)$~MHz, respectively. At $L=4~\mu$m, there are four $|\mu_{\beta}\rangle$ with $\Delta E/h$ lower than 100~MHz, with $\lvert\langle ngng|\mu_{\beta}\rangle|^2=(210,4.7,0.013,0.0058)\times10^{-4}$ and $\Delta E_{\beta}/h=(71,71,5.8,75)$~MHz, respectively. At $L=3~\mu$m and $L=4~\mu$m, the average blockade are $\overline{\Delta E}/h= 1.9$~GHz and $0.81$~GHz, respectively. These show that the interaction is strong to preserve the blockade mechanism.

\subsection{Van der Waals interaction }\label{sec05B2}
Figure~\ref{figure-C3gf0} shows that with $L$ increasing, there is one curve with more and more overlap with $\lvert ngng\rangle$, which means that it is reasonable to use one parameter, the $C_6$ coefficient~\cite{Walker2008}, for describing the interaction. This is because when the distance between two Rydberg atoms is large so that the dipole-dipole interaction barely couples the initial two-atom Rydberg state to another two-atom Rydberg state, there will be an energy shift to the two-atom Rydberg state which can be characterized by the second-order perturbation theory~\cite{Shi2021qst}.

\begin{figure}
\includegraphics[width=3.4in]
{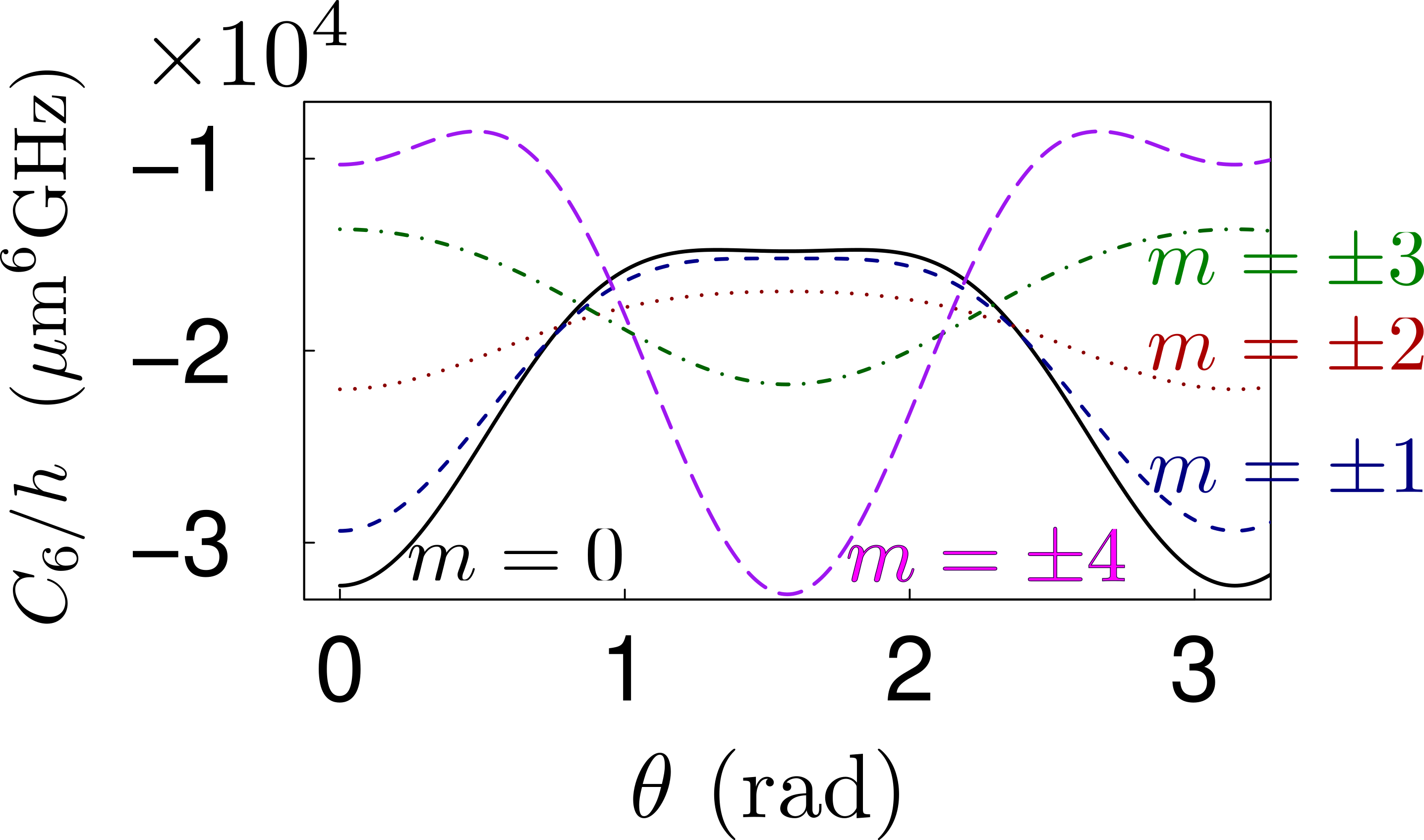}
\caption{ The van der Waals interaction coefficient $C_6$ in units of $h\times10^4\mu$m$^6$GHz for two atoms in the state $\lvert 6sng~^1G_4, m_j=m\rangle \otimes \lvert 6sng~^1G_4, m_j=m\rangle$ with $n=78$ as a function of the angle between the quantization axis and the two-atom separation direction. Among the cases shown here, the $C_6$ of the Rydberg state with $m=\pm4$ has the largest variation when $\theta$ varies, and the smallest $C_6$ here, about $-h\times8.6\mu$m$^6$THz, is at around $\theta\approx0.48$ or $5.81$~rad with $m=\pm4$.    \label{figure-C6-78} }
\end{figure}

By using the quantum defects suggested in Ref.~\cite{Neukammer1988}, the diagonal matrix element of the $C_6$ coefficient is shown in Fig.~\ref{figure-C6-78} with principal quantum number $n=78$. Here we choose $n=78$ because it is the highest $^1G_4$ barium state with the energy measured experimentally~\cite{Neukammer1988}. To examine whether the calculation is reasonable, we use the $C_6$ from Fig.~\ref{figure-C6-78} to plot $C_6/L^6$ in Fig.~\ref{figure-C3gf0}, which shows good overlap between the dashed curve and the emerging curve formed by the heaviest green colored dots at large $L$. To zoom in, we take $L=10~\mu$m as an example, and find that in the data from Hamiltonian diagonanization, the two $\lvert \mu_\beta\rangle$ most overlapped with $\lvert ngng\rangle$ has $\Delta E_\beta/h=(-9.2,~-9.3)$~MHz and $|\langle ngng|\mu_\beta\rangle|^2=(0.47, 0.48)$ for $m_j=2$, $\Delta E_\beta/h=(-7.1,~-8.6)$~MHz and $|\langle ngng|\mu_\beta\rangle|^2=(0.34, 0.48)$ for $m_j=3$, and $\Delta E_\beta/h=(290,~-10)$~MHz and $|\langle ngng|\mu_\beta\rangle|^2=(0.023, 0.96)$ for $m_j=4$, respectively. On the other hand, the value of $C_6/L^6$ is $h\times(-22, -14,~-10)$~MHz with $m_j=(2,~3,~4)$ if using the $C_6$ of Fig.~\ref{figure-C6-78}. These data mean that at $L=10~\mu$m the $m_j=4$ state has an interaction that can be well captured by the $C_6/L^6$ scaling, while for $m_j=2$ and 3, the interaction from the perturbation theory is about twice of the result from Hamiltonian diagonanization. This is possibly due to that for states with smaller $|m_j|$, too many states are coupled, which slows down the convergence of the dipole-dipole interaction toward a simple $C_6/L^6$ scaling.

Compared to alkali-metal atoms, the interaction of $6sng~^1G_4$ barium state is strong. For example, for two atoms each of which is in the state $\lvert 78s^2S_{1/2}, m=1/2\rangle$ of rubidium or cesium, we would have $C_6/h\approx3.1~\mu$m$^6$THz or $C_6/h\approx2.4~\mu$m$^6$THz, which is about three or four times smaller than the smallest $C_6$ of barium, i.e., $C_6/h\approx-10~\mu$m$^6$THz with $|m_j|=4$ in Fig.~\ref{figure-C6-78}. This means that the $6sng~^1G_4$ barium Rydberg states can provide strong interaction for the Rydberg blockade mechanism~\cite{Saffman2010}.

It is also useful to compare the interaction of barium with other divalent atoms. In Ref.~\cite{Peper2025}, a careful examination was made on the Rydberg interaction of $^{171}$Yb. Reference~\cite{Peper2025} showed that $C_6/h=34~\mu$m$^6$GHz for the $^{171}$Yb $|6s\nu~^3S_1,F=1/2, m_F=1/2\rangle$ state with an effective principal quantum number $\nu = 54.28$ and $\theta=\pi/2$. For barium state, the $6s54g~^1G_4, m_j=4$ state has $\nu \approx 53.94$, which is near to the one of Ref.~\cite{Peper2025}, and we find that it has $C_6/h=-570~\mu$m$^6$GHz when $\theta=\pi/2$, which is over 16 times of that in Ref.~\cite{Peper2025}.

\section{Prospects on quantum nonlinear optics with barium}\label{polariton}
It was shown in Ref.~\cite{Paredes-Barato2014} that quantum information processing based on Rydberg polaritons can be realized based on Rydberg interactions. However, typical experiments involving Rydberg polaritons with rubidium or cesium showed that their $T_{2,\text{\tiny D}}$ is on the order of microsecond~\cite{Dudin2012,Peyronel2012,Maxwell2013,Tiarks2018,PhysRevLett.131.133001}. These make it challenging to carry out high-fidelity photon-photon entangling gates based on Rydberg polaritons with cesium or rubidium. Below, we show that it is possible to explore quantum nonlinear optics based on barium-based Rydberg polaritons with long $T_{2,\text{\tiny D}}$.

\subsection{Rydberg polaritons with long $T_{2,\text{\tiny D}}$  }\label{polariton1}
The transition in Eq.~(\ref{Ba-Rydberg}) form a ladder-type three-level system which is useful for realizing quantum nonlinear optics with Rydberg polaritons~\cite{Dudin2012,Peyronel2012,Firstenberg2013,Baur2014,Tiarks2014,Gorniaczyk2014,Tiarks2016,Tiarks2018,PhysRevLett.127.063604,Jiao2024}. In these applications, a single photon can be loaded into an atomic medium as the gate photon, then sending another photon as the target photon through results in an interaction effect. If there is no Rydberg interaction, the target photon can propagate through without absorption when the detuning $\Delta_{^1\text{F}_3}$ at $5d6p~^1F_3^{\text{o}}$ is zero. But because of Rydberg interaction, the target photon can be blocked, experiencing loss, or even be completely scattered when $\Delta_{^1\text{F}_3}=0$ and the Rydberg interaction is strong~\cite{Gorshkov2013,Baur2014,Gorniaczyk2014,Tiarks2014}. When $\Delta_{^1\text{F}_3}$ is larger than the linewidth of $5d6p~^1F_3^{\text{o}}$, an interaction-induced phase can appear~\cite{Gorshkov2011,Tiarks2016,Tiarks2018}. In order to have a desired $\pi$ phase imprinted so as to realize a CNOT gate~\cite{Tiarks2018}, the optical depth of the atomic medium should be large, i.e., the density of the medium should be high. This causes slow group velocity for the target photon, so that the time for the target photon to fly through the atomic medium can be several microseconds~\cite{Tiarks2018}. For rubidium or cesium, the Doppler-limited coherence time of the stored gate photon in Eq.~(\ref{eq-T2}) is on the order of 20~$\mu$s with an atomic temperature $1~\mu$K as shown in Table~\ref{table-0}. Because of this, even though cooling atoms to a temperature of order of $1~\mu$K is very challenging, such cooling is necessary for carrying out deterministic photonic quantum information processing with alkali metal atoms. For instance, the atomic media were cooled to around $0.5~\mu$K in Refs.~\cite{Tiarks2016,Tiarks2018}.

The applications of Rydberg-mediated quantum nonlinear optics~\cite{Gorshkov2013,Baur2014,Gorniaczyk2014,Tiarks2014,Gorshkov2011,Tiarks2016,Tiarks2018} with barium can benefit from the long coherence time if one chooses the transition in Eq.~(\ref{Ba-Rydberg}). In Table~\ref{table-0}, one can see that for the same motional temperature in the atom gas, the motional coherence time of barium is about 47~(37) times that of rubidium~(cesium). For an operation time around $5~\mu$s with the CNOT gate of Ref.~\cite{Tiarks2018}, the motional coherence time of barium $T_{2,\text{\tiny D}}\sim 90~\mu$s with a relatively high atom temperature $T_{\text{eff}}=100~\mu$K
is still long enough to preserve the capability to achieve high fidelity. But if we choose rubidium to achieve $T_{2,\text{\tiny D}}\sim 90~\mu$s, we should cool atoms to $T_{\text{eff}}=0.05~\mu$K. Thus using the ladder-type transition in Eq.~(\ref{Ba-Rydberg}) for Rydberg-mediated quantum nonlinear optics has a potential advantage.

\subsection{$1.5~\mu$m-telecom-band quantum nonlinear optics with barium }\label{polariton2}
Quantum optics with photons in the telecommunication window are of practical interest because photons of wavelengths $1.25-1.65~\mu$m can be sent in fibers with low transmission loss. If we consider the 1500~nm transition between the metastable state $6s5d~^1D_2$ and the excited state $6s6p~^1P_1^{\text{o}}$, trapping barium atoms by optical cavities or silicon photonic crystal cavity can enable quantum networks with neutral atom processing nodes~\cite{Covey2019prappl}. Reference~\cite{Covey2019prappl} analyzed the 1389~nm transition $6s6p~^3P_1^{\text{o}}\leftrightarrow 6s6p~^3D_1$ of $^{171}$Yb for telecom-band quantum optics, which showed that a single-atom cooperativity $C_0=4g_0^2/\kappa\Gamma$ about 47 is feasible if the cavity finesse is $F=2000$, where $g_0$ is the coherent coupling rate between the atomic transition and the cavity field, $\kappa$ and $\Gamma$ are the cavity linewidth and the transition linewidth, respectively. Note that the transition linewidth $\Gamma$ is usually smaller than the natural atomic linewidth since the upper state can decay to multiple lower states. The dipole matrix element for the transition $6s5d~^1D_2\leftrightarrow 6s6p~^1P_1^{\text{o}}$ of barium is $0.527ae_0$~\cite{supple}, and the transition linewidth $\Gamma$ is $3.0\times10^5$~s$^{-1}$~\cite{Niggli1987}, while these two numbers for the setup of Ref.~\cite{Covey2019prappl} are $1.63ae_0$ and $2.0\times10^6$~s$^{-1}$, respectively. Suppose that similar trapping can occur for barium as in~\cite{Covey2019prappl} for ytterbium, then we can assume similar electric field strength and cavity decay rate if barium atoms are used in the setup analyzed in Ref.~\cite{Covey2019prappl}. In this case, we can have a $C_0$ which is 0.7 times the $C_0$ estimated in Ref.~\cite{Covey2019prappl}. The large cooperativity only means that the atomic transition is strongly coupled to the cavity mode~\cite{Reiserer2015}, but for barium in free space the state $6s6p~^1P_1^{\text{o}}$ has a short lifetime $8.2$~ns~\cite{Dzuba2007}, which has a branching fraction equal to 0.9966, 0.0025, and 0.0009 for the decay to $6s^2~^1S_0, 6s5d~^1D_2$, and $6s5d~^3D_2$, respectively~\cite{Niggli1987}. Thus, cavities with large finesses are required to have coherent coupling between the atom and cavity mode when using the 1500~nm transition $6s5d~^1D_2\leftrightarrow 6s6p~^1P_1^{\text{o}}$ of barium for cavity-assisted telecom-band quantum nonlinear optics.

It is also possible to use barium for frequency transduction between optical and telecom-band photons. In Ref.~\cite{Son2024}, transduction of 1500-nm photons to 553-nm photons at room temperature using barium atoms via the three-level transition
\begin{eqnarray}
 6s^2~^1S_{0} \xleftrightarrow[ ]{ }  6s6p~^1P_1^{\text{o}}\xleftrightarrow[ ]{  } 6s5d~^1D_2\label{son2024transduction}
\end{eqnarray}
was experimentally demonstrated. The essence of the transduction in Ref.~\cite{Son2024} is by preparing atoms in $6s5d~^1D_2$, an incoming 1500-nm photon will trigger a spontaneously emitted optical photon via the decay to $6s^2~^1S_{0}$, and atoms in $6s^2~^1S_{0}$ can be detected by using probe light resonant for the transition $6s^2~^1S_{0} \xleftrightarrow[ ]{ }  6s6p~^1P_1^{\text{o}}$~\cite{Son2024}.

\section{discussions}\label{sec12}
The barium atom has two valence electrons, which is similar to ytterbium, for which simultaneous trapping of the ground and high-n Rydberg states were demonstrated~\cite{PhysRevLett.128.033201}. The reason for this possibility is that the polarizability for the Rydberg state consists of two parts, that of the inner valence electron, and that of the outer Rydberg electron. In particular, for Rydberg state of large $n$ and trap waist of order of $1~\mu$m, there is little overlap between the Rydberg electron and the trap light~\cite{PhysRevLett.128.033201}. Then, the polarizability of the Rydberg atom is mainly from that of the inner valence electron. Therefore, it is possible to simultaneously trap both the ground state and the Rydberg state with an optical tweezer of equal depth for both states.

For the purpose of trapping the $ 6sng~^1G_4$ barium state, the concern of such feasibility is whether the polarizability of the ion core can be large enough. To analyze this, we note that for trapping ytterbium Rydberg states in Ref.~\cite{PhysRevLett.128.033201}, the polarizability of the Yb$^+$ core is mainly from the transitions $6s\rightarrow 6p_{1/2}$ and $6s\rightarrow 6p_{3/2}$, and the lifetime of Yb$^+6p_{1/2}$ is $8.12$~ns~\cite{PhysRevA.80.022502}. For $ 6sng~^1G_4$, the polarizability of the Ba$^+$ core arises also mainly from the transitions $6s\rightarrow 6p_{1/2}$ and $6s\rightarrow 6p_{3/2}$, where the lifetimes of Ba$^+6p_{1/2}$ and Ba$^+6p_{3/2}$ are $7.9$~ns and $6.32$~ns\cite{EHPinnington1995}, respectively. These data show that the transitions from the ground to the $6p_{1(3)/2}$ states can be a little stronger for barium compared to ytterbium, and, hence, it is possible to have a large polarizability for the $6s$ ion core. Therefore, a simultaneous trapping of the ground and $ 6sng~^1G_4$ Rydberg states should be possible. Further, SM~\cite{supple} shows that there are several magic wavelengths in the range $[665, ~1634]$~nm for trapping the two qubit states, therefore simultaneous trapping of the ground and $ 6sng~^1G_4$ Rydberg states is possible.

Since the trapping of high-n $ 6sng~^1G_4$ Rydberg states is mainly due to the polarizability of the inner valence electron, it is in principle possible to find triple-magic trapping condition for the two qubit states and the high-n $ 6sng~^1G_4$ Rydberg state. In Ref.~\cite{ammenwerth2024}, simultaneous trapping of the ground and a pair of metastable low-lying states of $^{88}$Sr was experimentally realized via fine tuning of wavelength and polarization of the trapping light. In theory, it is even possible to trap four states simultaneously~\cite{bhowmik2024double}. Therefore, simultaneous trapping of the $ 6sng~^1G_4$, the two qubit states, and a fourth state such as $6s5d~^3D_3$ may also be feasible. Then, the state $6s5d~^3D_3$ for shelving the qubit state $|1\rangle$, as studied following Eq.~(\ref{shelving01}) can be trapped together with the two qubit states and the Rydberg state, enabling prospects of higher-fidelity control for the barium-based Rydberg-atom technologies.

\section{conclusion}\label{sec13}
We have shown that by using barium-138 for defining qubits with a metastable d-orbital state, it is possible to realize a two-photon Rydberg excitation with a small motion-induced Rydberg Doppler dephasing rate, which is about 47~(37) times smaller than the smallest Rydberg dephasing rate in a two-photon Rydberg excitation of rubidium~(cesium) at a similar motional temperature. Moreover, the excited barium Rydberg states possess long radiative lifetimes and strong dipole-dipole interactions. These strengths mean that it is possible to use barium atoms for realizing quantum entangling gates with high fidelity. They also bring opportunities to realize all-optical quantum information processing which has been hindered by the fast Rydberg dephasing.

\section*{acknowledgments}
The research leading to the results here has received funding from the National Natural Science Foundation of China under Grants No. 12074300 and the Innovation Program for Quantum Science and Technology 2021ZD0302100. The author thanks the anonymous referees for their valuable comments, Yan Lu for useful discussions, and the Beijing Super Cloud Center for providing HPC resources that have contributed to the calculation of the Rydberg interactions of barium atoms.

%

\end{document}